# Noncollinear ferroelectric and screw-type antiferroelectric phases in a metal-free hybrid molecular crystal

Na Wang[1,‡], Zhong Shen[2,‡], Wang Luo[1], Hua-Kai Li[1], Ze-Jiang Xu[1], Chao Shi[1], Heng-Yun Ye[1,✉], Shuai Dong[2,✉], & Le-Ping Miao[1,✉]

[1]*Chaotic Matter Science Research Center, Department of Materials, Metallurgy and Chemistry, Jiangxi University of Science and Technology, Ganzhou 341000, China.*

[2]*Key Laboratory of Quantum Materials and Devices of Ministry of Education, School of Physics, Southeast University, Nanjing 211189, China.*

[‡]These authors contributed equally: Na Wang, Zhong Shen.

[✉]e-mails: (H.Y.Y.) hyye@seu.edu.cn, (S.D.); sdong@seu.edu.cn, (L.P.M.); miaoleping@jxust.edu.cn

## ABSTRACT

Noncollinear dipole textures greatly extend the scientific merits and application perspective of ferroic materials. In fact, noncollinear spin textures have been well recognized as one of the core issues of condensed matter, e.g. cycloidal/conical magnets with multiferroicity and magnetic skyrmions with topological properties. However, the counterparts in electrical polarized materials are less studied and thus urgently needed, since electric dipoles are usually aligned collinearly in most ferroelectrics/antiferroelectrics. Molecular crystals with electric dipoles provide a rich ore to explore the noncollinear polarity. Here we report an organic salt $(H_2Dabco)BrClO_4$ ($H_2Dabco$ = $N,N'$-1,4-diazabicyclo[2.2.2]octonium) that shows a transition between the ferroelectric and antiferroelectric phases. Based on experimental characterizations and *ab initio* calculations, it is found that its electric dipoles present nontrivial noncollinear textures with 60°-twisting angle between the neighbours. Then the ferroelectric-antiferroelectric transition can be understood as the coding of twisting angle sequence. Our study reveals the unique science of noncollinear electric polarity.

# INTRODUCTION

The scientific values of noncollinear magnetic orders in crystals have been gradually recognized in recent years, which have intimate connections with many important physical phenomena, such as magnetoelectricity in multiferroics[1], topological quasi-particles like skyrmions/bimerons[2], and (quantum) anomalous Hall effect[3]. These emergent effects provide broad materials to construct energy-efficient and high-speed information devices, e.g. the magnetoelectric spin-orbit (MESO) device[4].

The mapping relations between magnetism and polarity, e.g. (anti-)ferromagnetism *vs* (anti-)ferroelectricity and magnetization hysteresis *vs* polarization hysteresis, have been widely believed as one fundamental principle of matter, at least in the phenomenological level according to the Laudau theory[5–10]. Therefore, it is natural to expect entities to host similar noncollinear electric dipole orders and their associated emergent properties. However, the studies of noncollinear electric dipoles are much lagged compared with their magnetic counterparts[11–14], due to the following reasons. First, in ferroelectrics, the preferred polar axes are relative less, due to the locking of polarization with crystalline axes. Thus, the most possible cases between nearest-neighboring dipoles are parallel or antiparallel, except at the domain wall. In contrast, in magnets, the spin orientations are more free, slightly constrained by crystalline axes due to the weak spin-orbit coupling. Thus, the intrinsic noncollinear dipole textures are much less than the magnetic counterparts. Second, for many transitions which can lead to multiple polarization axes, the collective ferroelastic distortion does not prefer the noncollinear polarization. For example, for the *m*3*m*F4*mm* transition in $BaTiO_3$, the inequality between the *c*-axis in the polar direction and the *a*-axis in the non-polar direction does not allow adjacent unit cells to have polarization perpendicular to each other, for the high penalty of distortion energy. In fact, for most ferroelectrics with multiple polar axes, no anti-parallel antiferroelectricity has been observed, let alone noncollinear antiferroelectricity[15–18]. Thus, considering their scientific significance and implication on applications, noncollinear dipole orders are highly nontrivial and much anticipated.

Inspired by the idea of exchange frustration in magnets, it was proposed that the competition between ferroelectric and antiferroelectric phonon modes can lead to noncollinear canting order in $WO_2Cl_2/MoO_2Br_2$ monolayers[19]. Later and independently, a helical dipole order was detected in $BiCu_xMn_{7-x}O_{12}$[20], and a canting dipole order was found in La-doped $Pb(Zr,Sn,Ti)O_3$[21], with the this idea of interaction competition as underlying driving force. Besides this atomic level noncollinearity, mesoscopic scale polar vortices and polar skyrmions were predicted and realized in some artificial structures[22–24], imposed by the electrostatic boundary conditions. Despite these pioneering progresses, intrinsic noncollinear polarity remains rare in inorganic compounds, which limits the elbow room of their applications.

Alternatively, molecular-based organic-inorganic hybrid dielectrics/ferroelectrics provide a unique platform to pursue noncollinear dipole orders. Comparing with pure inorganic compounds, these hybrid molecular-based systems have more structural/element degrees of freedom for targeted design via chemical approaches, to pursue novel functions and optimize their performances[25–28]. These materials have advantages in high flexibility, lightweight nature, easy solution and vacuum processes, and easy film preparation. More importantly, they exhibit environmental friendliness and biocompatibility, and thus show advantages in biocompatible fabrication and applications, such as wearable electronics and bionic products[29–33].

Here, we report a hybrid molecular crystal $(H_2Dabco)BrClO_4$ (DBC, $H_2Dabco$ = $N,N'$-1,4-diazabicyclo[2.2.2]octonium) showing sequential ferroelectric-antiferroelectric-paraelectric phase transitions. More interestingly, intrinsic noncollinear dipole orders, i.e. a 60°-twisting angle between neighbour layers, are found in both the ferroelectric and antiferroelectric phases, while these two kinds of polarity can be distinguished by the twisting sequences. In this sense, its polarity is uniquely determined by the coding modes of noncollinearity.

**RESULTS**

**Crystal structures of DBC**

Dabco (1,4-diazabicyclo[2.2.2]octane) owns $D_{3h}$ point symmetry and (quasi)spherical space shape. Dabco can be easily protonated to form rich hydrogen bonding structures with the surrounding anions, such as chain-like structures. The characteristic spherical shape gives the nature of easy reorientation due to the low movement energy barrier, meanwhile, inducing the crystallization of high symmetry structures, like the trigonal/hexagonal/cubic[28,29]. Here are some ,examples: $H_2DabcoSiF_6$ (cubic $Fm\bar{3}m$)[34], $H_2DabcoNH_4(ClO_4)_3$ (cubic $Pa\bar{3}$)[35], $H_2DabcoH_3O(ClO_4)_3$ (cubic $P2_13$), and $H_2DabcoH_3O(BF_4)_3$ (hexagonal $P6_3/m$)[36]. Thus, Dabco and its derivatives are promising components for constructing high-symmetry ferroelectrics[29,31].

As shown in Fig. 1a, here the DBC is just an example of self-assembly of (quasi)spherical components: [$H_2Dabco$], [$ClO_4$], and Br. The cm-scale colourless single crystals of DBC were obtained using the slow evaporation method from a clear aqueous solution at room temperature (Fig. 1b). The phase purity of DBC was verified by the powder X-ray diffraction measurement (Supplementary Fig. 1). To monitor the phase transitions of DBC, the differential scanning calorimetry (DSC) analysis was performed on its powder samples. According to the two pairs of endothermic/exothermic peaks of DSC curves, DBC undergoes two reversible phase transitions around 348/333 K ($T_{C1}$) and 490/482 K ($T_{C2}$) (Supplementary Fig. 2).

To further understand these phase transitions, the crystal structures of DBC at 273 K, 365 K, and 500 K were determined by single-crystal X-ray diffraction. The low-temperature phase (LTP) at 273 K crystallizes in the orthogonal structure (space group No. 33 $Pn2_1a$, polar point group $mm2$) with $a$ = 26.8863(1) Å, $b$ = 7.7325(4) Å, $c$ = 10.0958(5) Å (Supplementary Table 1). Microscopically, the inorganic [$ClO_4$] and organic Br⋯[$H_2Dabco$] are stacking layer by layer along the $c$-axis, as shown in Fig. 1c.

In the $ab$-plane, [$H_2Dabco$] and Br ions form one-dimensional N-H⋯Br hydrogen bonded chains (Supplementary Fig. 3). Within the $ab$-plane layer, the orientations of all are parallel, but nontrivially they have a 60°-twist between neighbour layers along the $c$-axis, as shown in Fig. 1d (also Supplementary Fig. 4). Consequently, the structural asymmetry of these hydrogen bonded chains along the $b$-axis can be clearly visible in Fig. 1d &1e, implying the polar axis, as expected from its point group. Thus the ferroelectricity of DBC is expectable to originate from the Br⋯[$H_2Dabco$] chain dipoles, which will be directly verified in the following.

The [ClO$_4$] ions are sandwiched between layers of Br⋯[H$_2$Dabco] chains, as shown in Fig. 1c. In the *ab*-plane, the orientations of [ClO$_4$] ions form a special bi-stripy (i.e. up-up-down-down) sequence, which affects the neighbour N-H⋯Br hydrogen bonds by generating weak disproportion (Supplementary Fig. 3). Along the *c*-axis, the 60°-twist also appears in [ClO$_4$] ions (Supplementary Fig. 5).

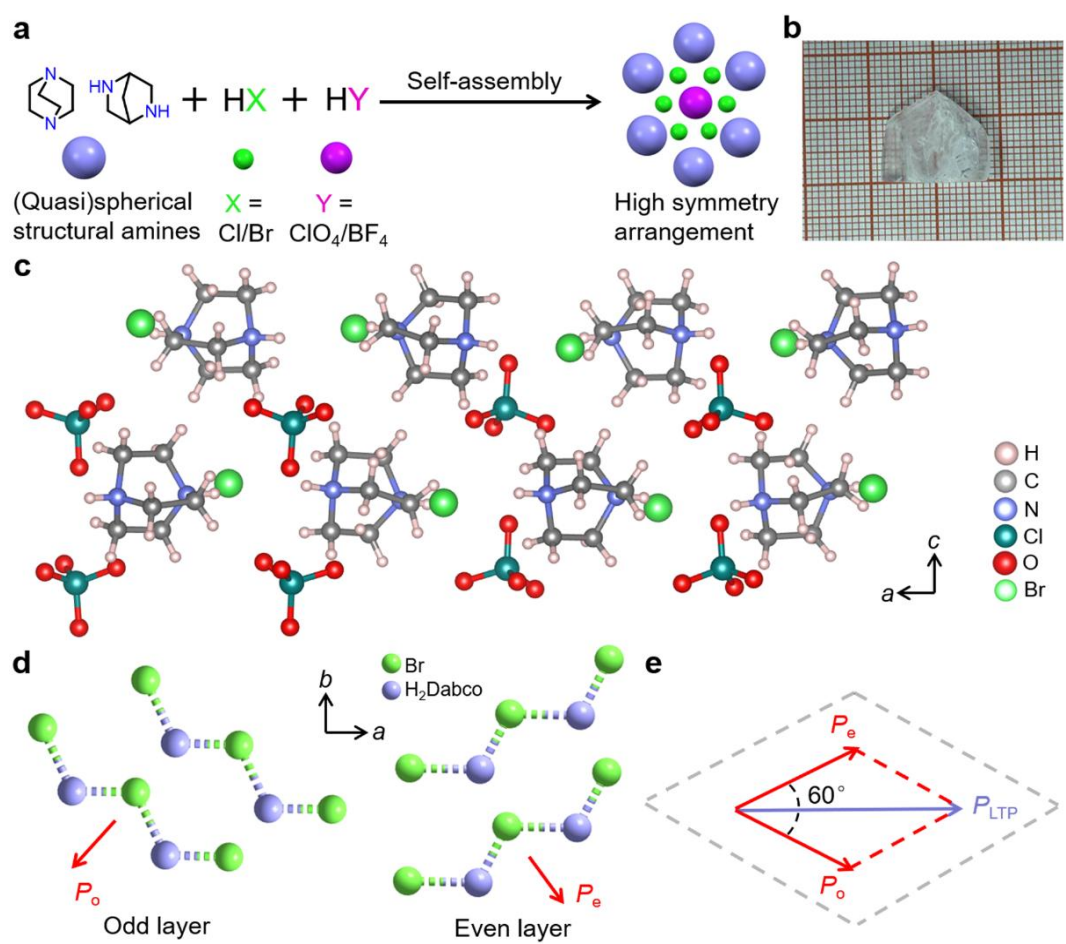

**Fig. 1 | Structures of DBC in LTP**. **a**, The self-assembly for the high symmetric arrangement of hybrid molecular DBC. **b**, The picture of cm-scale single crystal of DBC. **c**, The layered structure of [ClO$_4$] and Br⋯[H$_2$Dabco] in LTP. Along the *a*-axis, [ClO$_4$] ions form the up-up-down-down orientation model. More details of structures (LTP, ITP, and HTP) are shown in Supplementary Figs. 3-7. **d**, The packing of Br⋯[H$_2$Dabco] chains in LTP, viewed along the *c*-axis. Only the skeletons of chains are shown, while auxiliary ions are omitted for clarity. Between neighbouring layers, the orientations of Br⋯[H$_2$Dabco] chains (and associated electric dipoles *P*) are noncollinear, with 60° twist angle. **e**, The vector superposition of dipoles of even and odd layers ($P_e$ & $P_o$) in LTP.

For the intermediate-temperature phase (ITP), DBC adopts a higher-symmetry structure (space group No. 178 $P6_122$), which is nonpolar but chiral according to its point group 622. Microscopically, the oxygen ions in the [ClO$_4$] ions become disordered, i.e. with partial occupations on equivalent eight sites (Supplementary Fig. 6), no longer the up-up-down-down mode. Consequently, the weak disproportion of H···Br hydrogen bonds disappears, leading to a uniform value of 2.604 Å (Supplementary Fig. 3). The orientations of Br···[H$_2$Dabco] chains remain noncollinear between neighbour layers, i.e. with a ~60º twist around the *c*-axis. However, the twist mode changes from the stagger one (with a unit of 0º-60º twist) in the LTP to a screw one (with a unit of 0º-60º-120º-180º-240º-300º twist). Therefore, the *c*-axis is tripled (Supplementary Fig. 7). Such tripled structure cancels the net polarization from asymmetric [H$_2$Dabco] groups, by forming a rare screw antiferroelectric order of local dipoles. Meanwhile, these structural changes reduce the unit cell in the *ab*-plane to a quarter of the original one, from an orthogonal one to a rhombohedral one.

In the high-temperature phase (HTP), DBC has a centrosymmetric hexagonal structure (space group No. 194 $P6_3/mmc$, point group 6/*mmm*). Both [H$_2$Dabco] and [ClO$_4$] are disordered, i.e. with partial occupation, as shown in Supplementary Fig. 7. Consequently, the *c*-axis is further reduced to 1/3, without the orientation-twist of Br···[H$_2$Dabco] chains anymore. Therefore, the two phase transitions observed in DSC measurements correspond to the structural transitions from the ferroelectric LTP to the chiral antiferroelectric ITP at $T_{C1}$, and finally to the paraelectric HTP at $T_{C2}$.

**Experimental proof of ferroelectricity & antiferroelectricity**

Above structural analysis has revealed novel phase transitions involving possible ferroelectricity and antiferroelectricity with noncollinear orders of dipoles[20]. To verify these issues, more direct experimental characterizations were performed.

First, the variable-temperature second harmonic generation (SHG) was measured on powder samples. As shown in Fig. 2a, with increasing temperature, the SHG intensity drops quickly at $T_{C1}$. The strong SHG response below $T_{C1}$ is consistent with the ferroelectric polar structure. According to the Kleinman approximation, crystals with point group 622 should be

SHG-inactive. Accordingly, the SHG intensity above $T_{C1}$ is at the noise level.

Besides, the polar-nonpolar phase transitions can also be traced by the dielectric measurements[37]. As shown in Fig. 2b & Supplementary Fig. 8, the dielectric anomaly of DBC shows significant anisotropy along the three axes. Namely, the dielectric constant along the *b*-axis ($\varepsilon'_b$) is much larger than that along the *a*-axis ($\varepsilon'_a$), e.g. $\varepsilon'_b$ is ~10 times larger than $\varepsilon'_a$ at 1 MHz around room temperature. Most significantly, the peak of $\varepsilon'_b$ is very strong, unambiguously indicating the ferroelectric transition at $T_{C1}$ and polar axis. Interestingly, there is also a relatively weak peak of $\varepsilon'_a$, although the *a*-axis is not the polar axis according to the point group. It should be attributed to the antiferroelectricity along the *a*-axis, which will be further discussed. The dielectric constant along the *c*-axis ($\varepsilon'_c$) is even weaker than $\varepsilon'_a$, and there is no discernable anomaly (Supplementary Fig. 8).

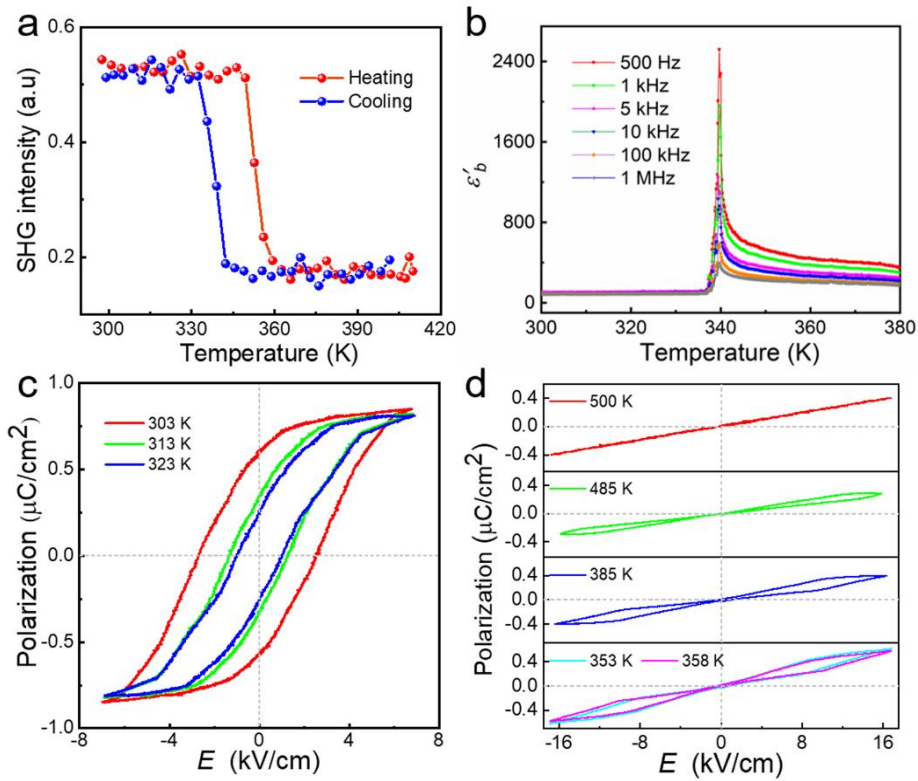

**Fig. 2 | Characterizations of ferroelectricity and antiferroelectricity of DBC crystal. a**, Evolution of SHG intensity upon heating and cooling (300-410 K). A sharp polar-nonpolar transition at $T_{C1}$ is obvious. **b**, The frequency-dependent dielectric constant along the *b*-axis upon cooling. Strong anomalies appear at $T_{C1}$ for all measured frequencies. **c**, The *P-E* hysteresis loops in the LTP region, exhibiting typical ferroelectric characteristics. **d**, The *P-E*

hysteresis loops in the ITP region, exhibiting typical antiferroelectric double-loop characteristics. The linear *P-E* curve of HTP is also presented. In **c** and **d**, the polarization is measured along the *b*-axis at 50 Hz.

In principle, the nonlinear polarization-electric field (*P-E*) hysteresis loop is the most persuasive proof of ferroelectricity and antiferroelectricity. As shown in Fig. 2c, the *P-E* hysteresis loops measured along the *b*-axis show typical ferroelectric characteristics below $T_{C1}$. The saturated polarization ($P_S$) is about 0.8 $\mu C/cm^2$ at 303 K. In contrast, the *P-E* hysteresis loop above $T_{C2}$ is fully linear, implying paraelectricity (Fig. 2d). The most interesting *P-E* behavior is observed between $T_{C1}$ and $T_{C2}$, as shown in Fig. 2d and Supplementary Fig. 9, which shows double hysteresis loops, a typical characteristic of antiferroelectricity. Based on the *P-E* loops, the saturated polarization in the ITP region is about 0.6 $\mu C/cm^2$, close but slightly lower than the LTP one, which is under expectation considering the temperature effects.

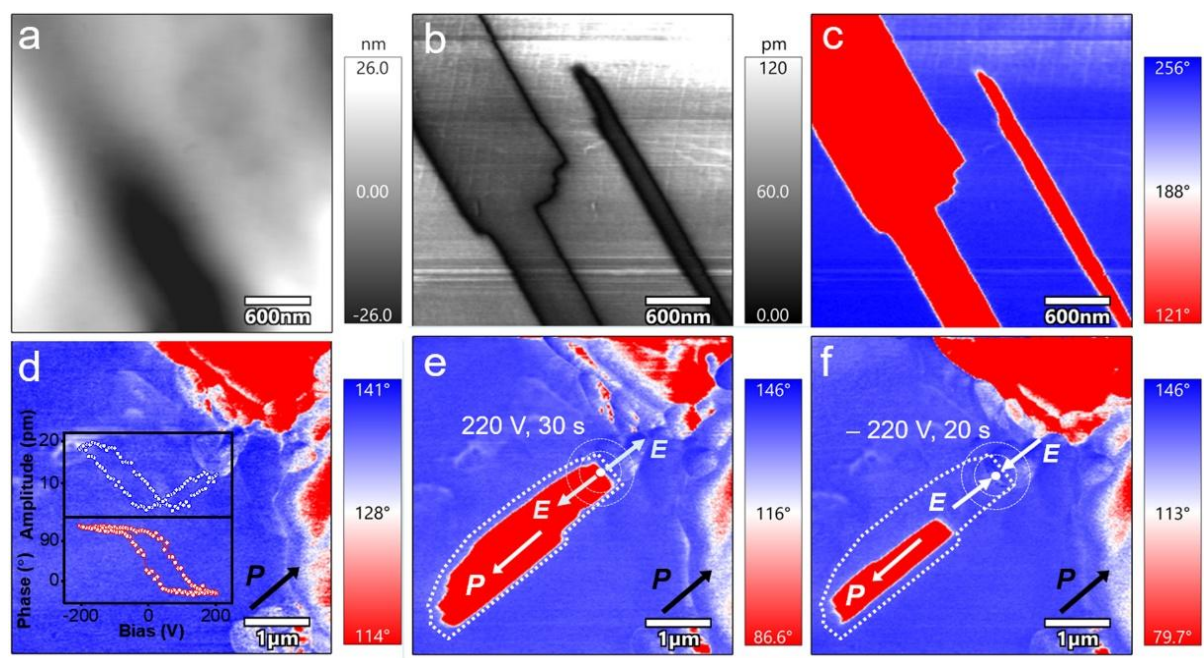

**Fig. 3 | PFM characterization of ferroelectric domains of DBC thin film. a**, The vertical PFM morphology image. **b**, The corresponding vertical PFM amplitude image. **c**, The corresponding vertical PFM phase image. **d**, The initial vertical PFM phase images. Inset: phase-voltage hysteresis loop and amplitude-voltage butterfly loop. **e**, Domain pattern after applying a positive bias of 220 V for 30 s at the marked area. **f**, Domain pattern after applying

a negative bias of 220 V for 20 s at the marked area.

The ferroelectric domain structure can be visualized and manipulated by piezoresponse force microscopy (PFM), which is another convincing evidence for ferroelectricity[38]. Figure 3a shows the topography of DBC film coated on indium-tin-oxide (ITO) substrate. Both the amplitude and phase images in vertical PFM clearly display domain structures with different polarization directions, where the domain walls appear as darker lines. (Figs. 3b & 3c). To further verify the ferroelectric switching behaviour, we conducted the point-wise polarization switching tests and recorded a vertical PFM switch spectroscopy loop. As shown in Fig. 3d, the typical hysteresis loop and butterfly shape curve are solid evidences for the polarization switch.

In addition, the ferroelectric domain switching process can be visualized through the PFM tip poling experiments. For example, by applying a positive bias of +220 V on a polarization-down position for 30 s, the domain can be completely reversed and the domain wall can be moved correspondingly, as shown in Fig. 3e & Supplementary Fig. 10. Moreover, by applying a tip bias of -220 V on the same position for 20 s can switch the polarization back, as shown in Fig. 3f & Supplementary Fig. 10. In addition, the temperature-dependent ferroelectric domain evolution is also investigated (Supplementary Fig. 11). The ferroelectric domain structures of DBC are visible in LTP. When heating to ITP, the domain structures disappear completely. While back to LTP, domain structures reappear and are significantly different from the initial ones. All these evidences unambiguously confirm the switchable nature of its ferroelectric polarization in LTP.

**DFT calculation of ferroelectricity & antiferroelectricity**

To further clarify the physical mechanisms of ferroelectricity and antiferroelectricity observed in DBC, we performed first-principles density functional theory (DFT) calculations. First, the crystal structures of LTP and ITP are fully optimized and their theoretical lattice constants are close to the experimental ones, as summarized in Table 1. Both of them are highly insulating with DFT band gaps of 4.7 eV and 4.5 eV for LTP and ITP respectively, as demonstrated in Supplementary Figs. 12–14. Note the experimental band gap is 4.9 eV at room temperature (Supplementary Fig. 15), very close to (only slightly higher than) the DFT one. Second, the DFT energy of LTP is indeed lower than that of ITP for 58.7 meV/f.u., implying the ground state. Third, the calculated ferroelectric polarization is 2.0 μC/cm² along the *b*-axis. Compared with the experimental results $P_b$ = 0.8 μC/cm² at 303 K, the theoretical value (at 0 K) is in the same order of magnitude. Although the theoretical one is somehow larger, but it remains reasonable considering the temperature effects and possible insufficiently polarized domains during the measurements. To confirm this point, we also test the experimental LTP structure at 273 K without DFT optimization, which leads to 0.9 μC/cm², very close to the experimental one. In short, the ferroelectric ground state can be well understood in our DFT calculation.

**Table 1 | Comparisons of experimental (Exp) and DFT results (at 0 K) of DBC.**

|  | FE | | | AFE | | |
| --- | --- | --- | --- | --- | --- | --- |
|  | Exp (273 K) | DFT | Difference | Exp (365 K) | DFT | Difference |
| *a* (Å) | 26.8863 | 26.2789 | -2.3% | 7.7979 | 7.6674 | -1.7% |
| *b* (Å) | 7.7325 | 7.6049 | -1.7% | - | - |  |
| *c* (Å) | 10.0958 | 10.2790 | 1.8% | 30.280 | 30.5599 | 0.9% |
| *V* (Å³) | 2098.90 | 2054.23 | -2.1% | 1594.56 | 1555.87 | -2.4% |

**DISCUSSION**

As aforementioned in the structural analysis (Fig. 1d), the local dipoles can be characterized by the Br⋯[H₂Dabco] chains, whose orientations determine the polar axis as evidenced in the ferroelectric LTP. As shown in Fig. 4a, the Br sites form a honeycomb lattice when projected to the *ab*-plane, which is composed of an upper triangle and a lower triangle in the nearest-neighbor layers. These two triangles rotate by 60° around the *c*-axis, which is a direct driving force for the aforementioned 60°-twist angle of Br⋯[H₂Dabco] chains. For the ferroelectric LTP, the sequence of twisting angle is stagger: +60°/–60°, which cancels the polarization along the *a*-axis (i.e. antiferroelectric axis) but leads to a net polarization along the *b*-axis (i.e. ferroelectric axis). Strictly speaking, this LTP is not a plain ferroelectric one, but a noncollinear ferroelectric one as proposed in $WO_2Cl_2$/$MoO_2Br_2$[19].

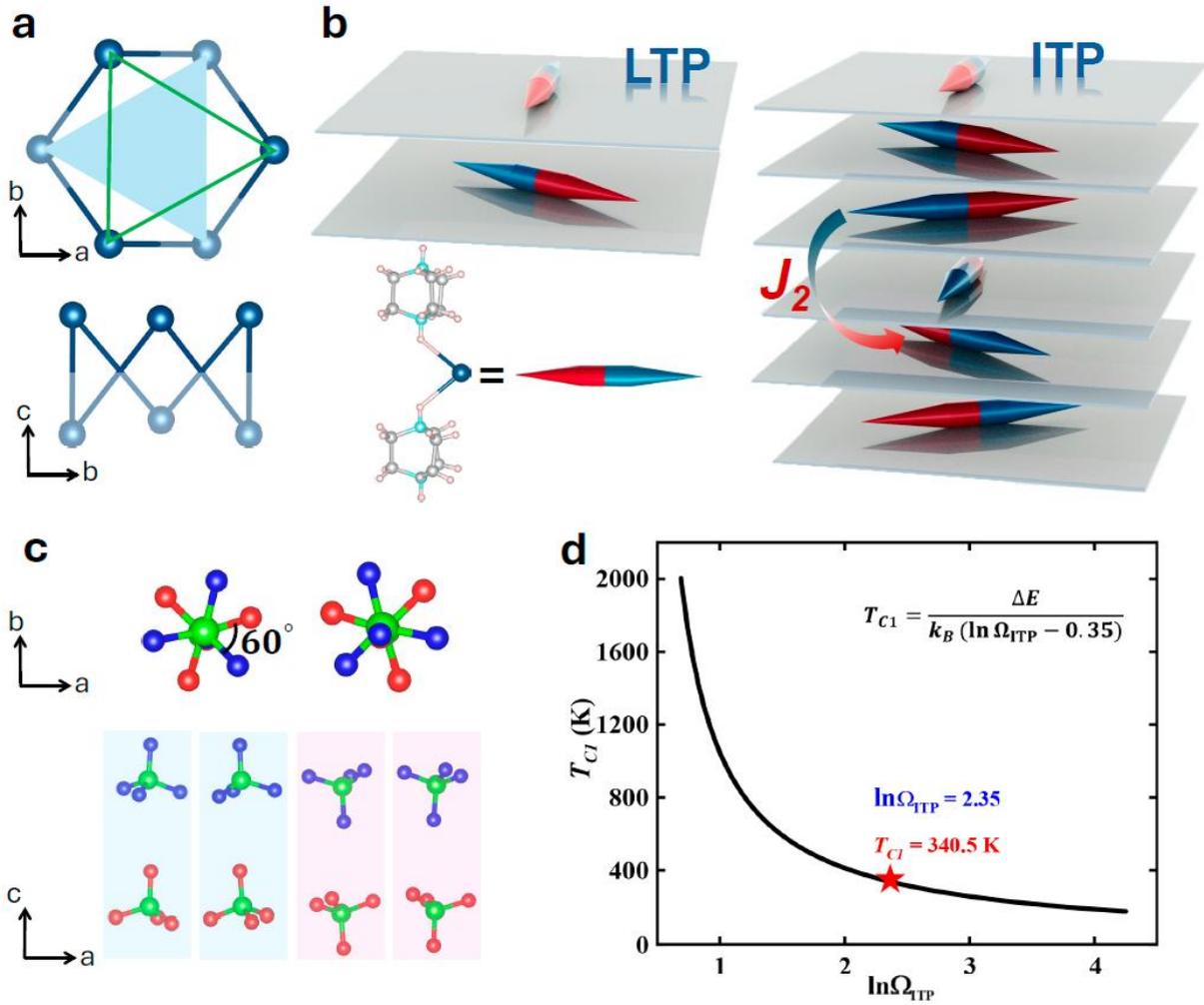

**Fig. 4 | Noncollinear polarity of DBC. a**, Top view of Br sublattice along the *c*-axis. In each layer Br ions forms triangles, and two neighbour layers of Br-triangles twist by 60°. **b**, Schematic diagram of noncollinear orders of LTP and ITP. The Br⋯[H$_2$Dabco] unit breaks the inversion symmetry, which can be represented by a dipole lying in the *ab*-plane. Locked by the Br-triangles, the dipoles twist by 60° between the nearest neighbour layers. The difference between LTP and ITP is the twisting sequence, which can be determined by the sign of the next-nearest neighbour interaction ($J_2$). **c**, The alignment of [ClO$_4$] groups in LTP. Here the different layers are distinguished by O's colours. Along the *c*-axis, the [ClO$_4$] anions also have the 60°-twist. Along the *a*-axis, the orientations of [ClO$_4$] anions form the up-up-down-down pattern. **d**, A rough estimation of $T_{C1}$ of entropy-driven LTP-ITP phase transition. The corresponding state number $\Omega_{ITP}$ is within the reasonable range.

Even though, the structural characteristics of [ClO$_4$] should also implicitly contribute to the formation of noncollinear dipole order. To clarify this point, we performed structural fine-tuning by partially replacing the components with similar geometries. (H$_2$Dabco)ClClO$_4$, (H$_2$Dabco)BrBF$_4$, and (H$_2$Dabch)BrClO$_4$ (Dabch = (*1S,4S*)-2,5-Diazabicyclo[2.2.1]Heptane) were synthesized. These compounds have the similar crystal packings to that of DBC. However, (H$_2$Dabco)ClClO$_4$, has a centrosymmetry space group. (H$_2$Dabco)BrBF$_4$ may undergo a ferroelectric phase transition but no noncollinear polarization was found. (H$_2$Dabch)BrClO$_4$ also has no polar symmetry. Therefore, [H$_2$Dabco], [ClO$_4$] and Br all contribute to the nontrivial noncollinear polarization and their contribution is synergic.

Then another interesting question is the origin of antiferroelectricity in the ITP. The key difference between ITP and LTP is the sequence of twisting angles along the *c*-axis: staggered for LTP but uniform for ITP. This difference may originate from the next-nearest neighbor interaction ($J_2$ as show in Fig. 4b). If such an interaction is ferroelectric preferred, the twist will be staggered, as in LTP. Else if this interaction prefers antiferroelectric alignment, the twist will be uniform, as in ITP. Since the next-nearest neighbor interaction is naturally weak considering the long distance, it is reasonable to be fluctuating around zero, whose sign can be fine tuned by surroundings such as the [ClO$_4$] ions. In the ITP, the [ClO$_4$] groups are in a disordered state, with eight partially occupied O-sites for each group (Supplementary Fig. 6). In contrast, in the LTP, the [ClO$_4$] ions are ordered. As aforementioned, the orientation of [ClO$_4$] groups is nontrivial, as shown in Fig. 4c. Along the *c*-axis, all [ClO$_4$] tetrahedra are almost parallelly aligned but with a 60°-rotation. But in the *ab*-plane, they form bi-strips with opposite alignment. According to the knowledge from magnetism, the bi-stripy texture can be obtained via certain exchange frustration. Similar mechanism can be applied here. Such a complex configuration of [ClO$_4$] tetrahedra may change the sign of $J_2$, making the stagger twist of Br⋯[H$_2$Dabco] chains to be more stable in LTP.

In addition to this idea based on polar frustration[19], another possible mechanism to drive noncollinear ferroelectrics and antiferroelectrics is the so-called electric Dzyaloshinskii-Moriya-like interaction (eDMI)[39]. Such a scenario based on eDMI needs a net polarization along the screwing axis to break the symmetry, as occurring in BiCu$_x$Mn$_{7-x}$O$_{12}$[20]. However, since here the ITP is nonpolar, the eDMI mechanism can be safely excluded.

Although the ITP owns a higher energy than LTP, its entropy (*S*) is higher, mainly due to the disorder of [ClO$_4$] ions. Therefore, its free energy *F* at finite temperatures may be lowered by the entropy, driving the LTP to ITP phase transition. Here we performed a rough estimation of entropy. The antiferroelectric ITP phase, with partially occupied O-sites in [ClO$_4$] ion, will lead to larger entropy due to more possible states. For each [ClO$_4$] ion, there are eight O-sites for four oxygen ions. Thus the possible states $\Omega_{ITP}$ is within the range 2 (for the strongly correlated limit) and 70 ($C_8^4$=70 in the totally random limit). Then the entropy per [ClO$_4$] ion $S_{ITP}$=$k_B$ln$\Omega_{ITP}$, is between 0.69$k_B$ and 4.25$k_B$, where $k_B$ is the Boltzmann constant (~1.38 × 10$^{-23}$ J/K). For the ferroelectric LTP phase, four [ClO$_4$]$^-$ groups has four possible states, due to the up-up-down-down (down-down-up-up and other variants) configurations. Then its entropy per [ClO$_4$] $S_{LTP}$ is $k_B$ln4/4=0.35$k_B$.

Then the $T_{C1}$ can be roughly estimated as $\Delta E/[k_B(S_{ITP}-S_{LTP})]$, where $\Delta E$ is the energy difference (~58.74 meV from DFT) per formula unit between LTP and ITP. According to the experimental $T_C$~340.5 K (the average value of the experimental $T_{C1}$ (348/333 K)), the effective entropy per [ClO$_4$] is 2.35$k_B$., which is in the middle of two limits 0.69$k_B$ and 4.25$k_B$, as sketched in Fig. 4d. Of course, such an estimation is only semiquantitative in a reasonable range, which should not be considered as a precise calculation.

Our study reveals an organic-inorganic hybrid metal-free molecular crystal DBC showing the ferroelectricity and antiferroelectricity at different temperatures. Interestingly, both the ferroelectric and antiferroelectric phases display intrinsic noncollinear dipole orders with a common 60°-twist between nearest-neighbor layers. The twisting mode between next-nearest-neighbor layers distinguishes the ferroelectric and antiferroelectric phases.

Our finding not only simply adds one candidate to the zoo of noncollinear ferroelectrics, but also provides an effective avenue to the rich ore of noncollinear polarity. One inborn advantage of organic-inorganic hybrid molecular dielectrics is their rich structural degrees of freedom, which allows various stereo orientations of local dipoles and leaves enough room to tune their noncollinearity. However, also due to the structural complexity, the underlying physical mechanisms were rarely clarified in hybrid molecular dielectrics previously. In this sense, this work is a seminal one to design and utilize noncollinear polarities for future applications.

Although the study of noncollinear ferroelectric materials remains in the early stage, the potential applications is promising. A noteworthy advantage of noncollinear dipole textures is that their switching barriers can be reduced in principle, due to their special switching paths via canting instead of flipping[19]. As compared in Supplementary Table 3, its coercive electric field can be relatively low. Thus the energy cost for ferroelectric switching will be saved, which is highly preferred for information devices, for example, the MESO devices[4]. Another potential usage is based on its chirality from additional degrees of freedom, which may be useful in electro-optic devices with electric-field controllable circular dichroism. More studies are needed in future to pursuit these applications.

**METHODS**

**Synthesis.** Crystal DBC was synthesized by mixing aqueous solutions of triethylenediamine (DABCO), $HClO_4$ (conc.), and HBr (conc.) in a molar ratio of 1:1:1. Massive colorless crystals were obtained by slow evaporation of solvent at room temperature after a few days.

**DSC, SHG, CD and XRD measurements.** Differential scanning calorimetry (DSC) measurements were performed on 214 Polyma instrument with a heating/cooling rate of 10 K/min under nitrogen. The SHG intensity was measured by pump Nd:YAG laser (1064 nm, 5 ns pulse duration, 1.6 MW peak power, 10 Hz repetition rate), and the temperature varies from 297 K to 410 K controlled by a precision temperature controller system (INSTEC Instruments, Model HCS302). The temperature-dependent (298, 303, 308, 363, 368, 373 K) CD spectrum upon the powder samples was measured on a circular dichroism spectrometer JASCO J-715 (Jasco, Tokyo, Japan). The Powder X-ray Diffraction were measured at a measurement angle of 5°-50° and a scan rate of 5°/min on Rigaku D/MAX 2000 PC X-Ray Diffractometer. Variable-temperature X-ray single-crystal diffraction analysis was carried out using a Rigaku synergy diffractometer with Mo-K$\alpha$ radiation ($\lambda$ = 0.71073 Å) radiation from a graphite monochromator. The crystal structures were solved by direct method and then refined by full-matrix least-square method based on $F^2$ using the OLEX2 and SHELXTL (2018) software package. All non-hydrogen atoms were refined anisotropically and located in different Fourier maps, and the positions of all hydrogen atoms were generated geometrically. The monocrystalline samples were oriented on the Rigaku synergy diffractometer (Operating system: CrysAlisPro 1.171.41.112a (Rigaku OD, 2021)) through the axial photographic functional module.

**Dielectric and ferroelectric measurements.** The single crystal direction of DBC was determined by the Rigaku synergy diffractometer [operating system: cryalispro 1.171.41.112a

(Rigaku OD, 2021)]. Subsequently, the single crystals were cut into thin plates (thickness ~0.4-0.6 mm) perpendicular to the crystalline *a*-, *b*-, and *c*-axes for dielectric and ferroelectric measurements. Complex dielectric permittivities were measured with a TH2828A impedance analyzer in the frequency range of 500 Hz to 1 MHz. The variable temperature *P-E* hysteresis loops of DBC were performed on the thin crystal through a Radiant Precision Premier II instrument with a frequency of 50 Hz. Ferroelectric switching measurements were directly carried out on the thin film by scanning probe microscopy technique through a resonant-enhanced PFM [MFP-3D, Asylum Research, and conductive Pt/Ircoated silicon probes (EFM-50, Nanoworld)].

**DFT calculation.** The first-principles density functional theory (DFT) calculations are carried out in the framework of Vienna *ab initio* Simulation Package (VASP)[40]. The generalized gradient approximation of Revised Perdew-Burke-Ernzerhof for solids (PBEsol) is used for the exchange-correlation functional[41]. The energy cutoff for the plane-wave basis is set as 520 eV. When optimizing the lattice constants and ionic positions, convergent criteria for energy and the Hellman-Feynman force are set as $10^{-6}$ eV and 0.03 eV/Å respectively. For the LTP and ITP phase, the Brillouin-zone integrations are carried out by using 1×3×5 and 5×5×1 G-centered *k*-point meshes, respectively. The polarization was calculated using the standard Berry phase method[42].

**Data availability**

All data generated and analyzed in this study are included in the article and Supplementary Information and are also available at the corresponding authors' request. The crystal structures generated in this study have been deposited in the Cambridge Crystallographic DataCenter under accession code CCDC: 2350391–2350393, 2388616–2388620 and can beobtained free of charge from the CCDC via www.ccdc.cam.ac.uk/data_request/cif.

**Acknowledgments**

L.-P.M. acknowledges support from National Natural Science Foundation of China (grant numbers 22205087 & 22371095) and Jiangxi Provincial Natural Science Foundation (grant numbers 20232BAB213003 & 20242BAB23015). S.D. acknowledges support from National Natural Science Foundation of China (grant numbers 12325401 & 12274069) and Big Data Center of Southeast University for providing the facility support on the numerical calculations. C. S. acknowledges the support from National Natural Science Foundation of China (Grant No. 22175079). H.-Y. Y. acknowledges the support from National Natural Science Foundation of China (Grant No. 21875093) and the Jiangxi Provincial Natural Science Foundation (grant number 20204BCJ22015).

**Author contributions**

L.P.M. and S.D. conceived the project. L.-P.M. & H.-Y.Y. designed the experiments. S.D. proposed the theoretical mechanisms. N.W. prepared the samples and performed the PFM measurements. Z.-J.X contributed to the *P-E* hysteresis loops measurement. W.L. & C.S. performed the DSC measurements. H.-K.Li contributed to single-crystal measurement and analysis. Z.S. performed the DFT calculations guided by S.D.. S.D. & L.P.M. wrote the manuscript, with inputs from all other authors.

**Competing interests:** The authors declare no competing interests.